\documentclass[twocolumn,showpacs,preprintnumbers,amsmath,amssymb]{revtex4}
\usepackage{graphics}
\usepackage{epsfig}
\usepackage{bm}

\begin{document}


\def\Tau{\mbox{\boldmath $\tau$}}

\title{\bf Thermodynamics of Ferrotoroidic Materials: Toroidocaloric Effect}

\author{Teresa Cast\'an and Antoni Planes}

\affiliation{Departament d'Estructura i Constituents de  la Mat\`eria, Facultat de  F\'\i sica,  Universitat de Barcelona. Diagonal, 647, E-08028
Barcelona, Catalonia, Spain}

\author{Avadh Saxena}

\affiliation {Theoretical Division, Los Alamos National Laboratory, Los Alamos, New Mexico 87545, USA}

\date{\today}

\begin{abstract}
The three primary ferroics, namely ferromagnets, ferroelectrics and ferroelastics exhibit corresponding large (or even giant) magnetocaloric,
electrocaloric and elastocaloric effects when a phase transition is induced by the application of an appropriate external field.  Recently the
suite of primary ferroics has been extended to include ferrotoroidic materials in which there is an ordering of toroidic moments in the form of
magnetic vortex-like structures, examples being LiCo(PO$_4$)$_3$ and Ba$_2$CoGe$_2$O$_7$. In the present work we formulate the thermodynamics of
ferrotoroidic materials.   Within a Landau free energy framework we calculate the toroidocaloric effect by quantifying isothermal entropy change (or
adiabatic temperature change) in the presence of an applied toroidic field when usual magnetization and polarization may also be present simultaneously.  We also obtain a nonlinear
Clausius-Clapeyron relation for phase coexistence.
\end{abstract}


\maketitle


\section{Introduction}
Ferroic phenomenon or the presence of switchable domain walls in an external field is a direct consequence of a specific broken symmetry
\cite{pt11}.  Loss of spatial inversion symmetry results in ferroelectricity whereas loss of time reversal symmetry results in ferromagnetism.
Ferroelasticity is a result of broken rotational symmetry although it remains invariant under both spatial inversion and time reversal symmetries.
The fourth possibility corresponds to when both spatial inversion and time reversal symmetries are simultaneously broken.  This is the case for
recently discovered ferrotoroidic materials \cite{schmid} where the long-range
order is related to an ordering of magnetic vortex-like structures  characterized by a toroidal dipolar moment. It is important to mention that the ferrotoroidic order is also related to magnetoelectric
behavior \cite{spaldin} which is one of the main attractions of multiferroics--materials that exhibit two or more ferroic orders simultaneously.
This class includes magnetoelastics as well as magnetic shape memory alloys.

In general, both electric and magnetic toroidal moments can be defined within the context of electromagnetism \cite{dubovik1990,dubovik2000}.
In the present paper we will exclusively consider the magnetic toroidal moment which is the only one that has the required spacial and time inversion
symmetries \cite{dubovik1990}.  We note that in recent years the study of the unique properties of ferroelectric materials
carrying an electric toroidal moment has received considerable attention \cite{prosandeev2007,prosandeev2008}

Ferrotoroidic domain walls have been observed in LiCo(PO$_4$)$_3$ using second
harmonic generation technique \cite{vanaken} and a symmetry classification of
such domains has been given recently \cite{litvin}.   Spontaneous toroidal
moments have been attributed to exist in the multiferroic phase of
Ba$_2$CoGe$_2$O$_7$ (BCG) \cite{toledano} and may be related to the observed
unusual magnetoelectric effects \cite{cheong, murakawa}.  Single crystalline
thin films of MnTiO$_3$ with an ilmenite structure also exhibit a
ferrotoroidic structure \cite{tokura}.  Neutron polarimetry indicates that the
magnetoelectric MnPS$_3$ is a viable candidate for ferrotoroidicity \cite{simonet}.
The magnetic phase transition in BiFeO$_3$ implies the appearance of a toroidal 
moment \cite{kadom} {\it Ab initio} calculations suggest that the olivine material 
Li$_4$MnFeCoNiP$_4$O$_{16}$ is possibly ferrotoroidic \cite{olivine}.  

A sequence of possibly two ferrotoroidal phase transitions has been considered 
phenomenologically for Ni-Br boracites \cite{sannikov}.  Another physical realization 
of toroidal order has been considered as an interacting system of discs with a 
triangle of spins on each disc \cite{harris}. Both charge and spin currents can lead 
to a toroidal state \cite{kopaev}.   Recent observation of orbital currents in CuO through 
resonant x-ray diffraction provides a direct evidence of antiferrotoroidic ordering 
\cite{scagnoli}   Ferrotoroidic materials exhibit linear magnetoelectric effect; in fact, 
the toroidic moment is related to an antisymmetric  component of the magnetoelectric 
tensor ($\alpha_{ij}\ne\alpha_{ji}$).   Moreover, the toroidic moment can be viewed as 
a quantum geometric phase \cite {batista}.  Beyond the magnetoelectric applications 
toroidic materials can also act as novel metamaterials \cite{meta}.

One way of realizing the physical consequences of toroidal moment is to characterize the thermal response of the material to an externally applied
field that couples to this moment and gives rise to measurable caloric effects. Actually, caloric effects are inherent to every material and are
commonly quantified by the adiabatic temperature change or to the isothermal entropy change that occur when an external field is applied or
removed. In the solid-state, the most studied caloric effect is the magnetocaloric effect \cite{review}, mainly after the discovery in the
mid-nineties of materials displaying giant magnetocaloric effect in the vicinity of room temperature \cite{Gschneidner2005}. However, in recent
years other caloric effects such as the electrocaloric \cite{Mischenko2006}, elastocaloric \cite{Bonnot2008} or barocaloric \cite{Manosa2010} have
also received considerable attention.

A crucial feature common to most materials exhibiting a giant caloric effect is the occurrence of a first-order phase transition. The expected
large (discontinuous) change of the order parameter at the transition involves a large entropy content (associated with the latent heat) which is
at the origin of the giant caloric effect. Moreover, a strong coupling between different degrees of freedom such as structural, magnetic,
electric, etc. enables the transition to be driven by different fields conjugated to such properties. The study of caloric effects is thus a
convenient method in order to study the thermodynamics of this class of complex materials. For instance, this should apply to multiferroic
materials which are expected to show more than one caloric effect, e.g. simultaneous electrocaloric and magnetocaloric effects, or more precisely
a magnetoelectrocaloric effect. The latter has not been reported yet.  The present paper deals with the thermal response and the associated
caloric effects resulting from changes of toroidal order in ferrotoroidic materials (undergoing a paratoroidic to ferrotoroidic transition). These
changes are likely to give rise to a toroidocaloric effect in this class of materials. Indeed, the study of toroidocaloric effect is expected
to provide new insights into understanding the important problem of switching the toroidal moment \cite{prosandeev2008}.

In this paper we study the thermodynamics of multiferroics and first derive
expressions for magnetoelectrocaloric and toroidocaloric equations in Sec. II
and Sec. III treating magnetization, polarization and toroidization as
independent order parameters.  In Sec. IV we present a Landau free energy for
a paratoroidic to ferrotoroidic transition with a specific coupling between
toroidization, polarization and magnetization, derive a phase coexistence
(Clausius-Clapeyron) relation and compute the toroidocaloric effect.
Possible experimental implications are discussed in Sec. V in which we also
present our main conclusions


\section{Caloric effects: general aspects}

Consider a generic thermodynamic system and take temperature ($T$) and generalized forces or fields ($\{\bf Y_i\}$) as independent variables.
Differential changes of the fields will yield a differential change of entropy [$S = S(T, \{{\bf Y_i} \})$] given by:
\begin{eqnarray}
dS & = & \frac{C}{T} dT + \sum_i \left(\frac{\partial S}{\partial {\bf Y_i}}\right)_{T, \{ {\bf Y_{j \neq i}} \}}  d{\bf Y_i} \nonumber \\
& = & \frac{C}{T} dT + \sum_i \left(\frac{\partial {\bf X_i}}{\partial T}\right)_{ \{ {\bf Y_j} \} }       d{\bf Y_i} \label{dS} ,
\end{eqnarray}
%
%
where $C/T = \left( \frac{\partial S}{\partial T} \right)_{\{{\bf Y_i}\}}$ defines the heat capacity at constant fields $\{{\bf Y_i}\}$ and we
have taken into account the general Maxwell relations:
\begin{equation}
\left(\frac{\partial S}{\partial {\bf Y_i}} \right)_{T, \{ {\bf Y_{j \neq i}} \}} = \left(\frac{\partial {\bf X_i}}{\partial T} \right)_{\{{\bf
Y_j}\}}  \label{Maxwell} .
\end{equation}
Here $\{{\bf X_i}\}$ denote generalized displacements thermodynamically
conjugated to the fields $\{{\bf Y_i}\}$. Any interplay between the different
degrees of freedom will be taken into account through the fact that any ${\bf
X_i}$ is, in principle, a function of all fields.  Therefore, the interplay
must be introduced through the state equations.

For each generalized displacement a caloric effect will occur when the
corresponding conjugated field is varied. Suppose, for instance that the field
${\bf Y_i}$ changes from 0 to ${\bf Y_i}$ so that the system passes from a
state $(T_i, 0)$ to $(T_f, {\bf Y_i})$, where $T_i$ and $T_f$ are the
temperatures of the initial and final states respectively. The corresponding
change of entropy is given by, $S(T_f, {\bf Y_i}) - S(T_i, 0)$. The two limits
of interest that quantify the caloric effect associated to the property ${\bf
X_i}$ conjugated to the field ${\bf Y_i}$, are the isothermal and the
adiabatic limits. In the isothermal case $T_i = T_f = T$ and the thermal
response is characterized by a change of entropy given by:
\begin{eqnarray}
\Delta S(T, 0 \rightarrow {\bf Y}) & = & S(T, {\bf Y_i}) - S (T, 0) \nonumber \\ & = & \int_0^{{\bf Y_i}} \left(\frac{{\partial \bf X_i}}{\partial
T}\right)_{{\{{\bf Y_j}\}}} d{\bf Y_i}.
\end{eqnarray}
In the adiabatic limit, $\Delta S = 0$, and the thermal response is quantified by the change of temperature given by:
\begin{eqnarray}
 \Delta T(T_{i}, 0 \rightarrow {\bf Y})  & = & T_f({\bf Y_i}) - T_i(0) \nonumber \\ & = &
 - \int_0^{{\bf Y_i}} \frac{T}{C}  \left(\frac{{\partial \bf X_i}}{\partial T}\right)_{{\{{\bf Y_j}\}}} d{\bf Y_i}.
\end{eqnarray}
Therefore, the caloric response of a material to a given field ${\bf Y_j}$ will be given by $\left(\frac{\partial {\bf X}_i}{\partial
T}\right)_{\{{\bf Y_j}\}}$.

\subsection{Examples}

In the case of the magnetocaloric effect the corresponding expression for the isothermal entropy change is:
\begin{equation}
\Delta S(T, 0 \rightarrow {\bf H}) = \int_0^{{\bf H}} \left(\frac{{\partial \bf M}}{\partial T}\right)_{{\{{\bf H}\}}} \cdot d{\bf H}
\end{equation}
and for the adiabatic temperature change it is:
\begin{equation}
\Delta T(T_{i}, 0 \rightarrow {\bf H}) = - \int_0^{{\bf H}} \frac{T}{C}  \left(\frac{{\partial \bf M}}{\partial T}\right)_{{\{{\bf H}\}}} \cdot
d{\bf H},
\end{equation}
where ${\bf H}$ is the magnetic field (strictly we should write $\mu_0 {\bf H} = {\bf B}$) and ${\bf M}$ is the magnetization. Notice that by just
replacing ${\bf H}$ by the electric field ${\bf E}$ and ${\bf M}$ by the polarization ${\bf P}$ the corresponding changes of entropy and
temperature that quantify the electrocaloric effect are obtained. Similarly, by replacing ${\bf H}$ by the stress ${\bf \sigma}$ and ${\bf M}$ by
the strain ${\bf \varepsilon}$ we get the corresponding changes of entropy and temperature for the mechanocaloric effect.  Therefore, a
barocaloric effect (a particular case of mechanocaloric effect) involving pressure $p$ and change in volume $\Delta V$ is also expected.

\section{Materials with toroidal order: Thermodynamics}

The three basic moments of the electromagnetic field are electric, magnetic and toroidal moments. Therefore, we assume that the three ferroic
properties are characterized by the corresponding moments per unit volume, polarization (${\bf P}$), magnetization (${\bf M}$) and toroidization
(${\bf \Tau}$) that can be assumed as independent (vector) order parameters. Here the toroidization is assumed to originate only from the
existence of magnetic toroidal moments. In fact, this appears to be the case for LiCo(PO$_4$)$_3$ \cite{vanaken} and Ba$_2$CoGe$_2$O
\cite{cheong, murakawa}. The corresponding thermodynamically conjugated fields will be the electric, ${\bf E}$, magnetic, ${\bf H}$, and toroidal,
${\bf G}$, fields. Therefore, the ``Thermodynamic Identity" for such a (closed) system reads:
\begin{equation}
dU = TdS + {\bf E}\cdot d{\bf P} + {\bf H}\cdot d{\bf M} + {\bf G}\cdot d{\bf \Tau} \label{ThermId} ,
\end{equation}
where $U$ is the internal energy per unit volume, $T$, temperature and $S$, the entropy per unit volume.  The field ${\bf G}$ coupling to the
toroidization is related to the electric and magnetic fields through  ${\bf G}
=  {\bf E} \times {\bf H}$. This choice, however, deserves some
discussion. Note that the natural conjugated field of the magnetic toroidal moment is ${\bf \nabla} \times {\bf B}$ \cite{dubovik1990}.
Nevertheless, since we are only considering homogeneous macroscopic bodies in thermodynamic equilibrium, from symmetry considerations we assume that
${\bf G}$ is the appropriate macroscopic field that enables external control of the toroidization. This is actually in agreement with references
\cite{spaldin} and \cite{dubovik1990} where it is shown that the free energy of a system with magnetic toroidal moment must include a term
proportional to the product ${\bf \Tau} \cdot {\bf G}$. It is worth pointing out that the field ${\bf G}$ cannot be modified independently of the
fields conjugated to polarization and magnetization. From the Thermodynamic Identity (\ref{ThermId}), this relationship between ${\bf G}$, ${\bf
E}$ and ${\bf H}$ requires that,
\begin{equation}
\left(\frac{\partial U}{\partial {\bf \Tau}} \right)_{S, {\bf P}, {\bf M}} = \left(\frac{\partial U}{\partial {\bf P}} \right)_{S, {\bf M}, {\bf
\Tau}} \times \left(\frac{\partial U}{\partial {\bf M}} \right)_{S, {\bf P}, {\bf \Tau}}.
\end{equation}
We now define the Gibbs free energy ${\cal G}$ through the following Legendre transform:
\begin{equation}
{\cal G} = U - TS  -{\bf E}\cdot {\bf P} - {\bf M}\cdot {\bf H} - {\bf \Tau} \cdot {\bf G}. \label{G}
\end{equation}
Differentiating this expression and replacing $dU$ in (\ref{G}), we obtain:
\begin{equation}
d{\cal G} = -S dT - [{\bf P} + ({\bf H} \times {\bf \Tau})] \cdot d{\bf E} - [{\bf M} + ({\bf \Tau} \times {\bf E})] \cdot d{\bf H}. \label{dG}
\end{equation}
Taking into account that double differentiation of $\cal{G}$ is independent of the order in which it is carried out, we obtain the following Maxwell
relations:
\begin{equation}
\left(\frac{\partial S}{\partial {\bf E}} \right)_{T,{\bf H}} = \left(\frac{\partial {\bf P}}{\partial T} \right)_{{\bf E},{\bf H}} + {\bf H}
\times \left(\frac{\partial {\bf \Tau}}{\partial T} \right)_{{\bf E},{\bf H}} \label{Max1}
\end{equation}
and
\begin{equation}
\left(\frac{\partial S}{\partial {\bf H}}\right)_{T,{\bf E}} = \left(\frac{\partial {\bf M}}{\partial T} \right)_{{\bf H},{\bf E}} +
\left(\frac{\partial {\bf \Tau}}{\partial T}\right)_{{\bf H},{\bf E}} \times {\bf E} \label{Max2} .
\end{equation}

We now take into account that polarization and magnetization can be written as
the sum of an intrinsic term originating from (pre-existing) free electric and
magnetic moments and a contribution arising from the toroidal moments. The
toroidal contributions satisfy \cite{pt11, spaldin, kopaev}:
\begin{equation}
{\bf P}_t = - {\bf \Tau} \times {\bf H} \label{tpolarization} ,
\end{equation}
and
\begin{equation}
{\bf M}_t = {\bf \Tau} \times {\bf E} \label{tmagnetization} .
\end{equation}
Then, the Maxwell relations (\ref{Max1}) and (\ref{Max2}) can be written in the form:
\begin{equation}
\left(\frac{\partial S}{\partial {\bf E}} \right)_{T,{\bf H}} = \left(\frac{\partial {\bf P}_i}{\partial T} \right)_{{\bf E},{\bf H}} +
\left(\frac{\partial {\bf P}_t}{\partial T} \right)_{{\bf E},{\bf H}} = \left(\frac{\partial {\bf P}}{\partial T} \right)_{{\bf E},{\bf H}}
\end{equation}
and
\begin{equation}
\left(\frac{\partial S}{\partial {\bf H}}\right)_{T,{\bf E}} = \left(\frac{\partial {\bf M}_i}{\partial T} \right)_{{\bf H},{\bf E}} +
\left(\frac{\partial {\bf M}_t}{\partial T}\right)_{{\bf H},{\bf E}} = \left(\frac{\partial {\bf M}}{\partial T}\right)_{{\bf H},{\bf E}} ,
\end{equation}
where the intrinsic contributions to the polarization and magnetization are related to the total as ${\bf P} = {\bf P_i}+{\bf P_t}$ and ${\bf M} =
{\bf M_i}+{\bf M_t}$.

A second set of Maxwell relations are obtained from differentiation with respect to ${\bf E}$ and ${\bf M}$. This yields:
\begin{equation}
\left(\frac{\partial {\bf P}_t}{\partial {\bf H}} \right)_{T, {\bf E}} = \left(\frac{\partial {\bf M}_t}{\partial {\bf E}} \right)_{T, {\bf H}} ,
\label{Max3}
\end{equation}
where we have taken into account that the intrinsic components ${\bf P}_i$ and
${\bf M}_i$ do not depend on ${\bf H}$ and ${\bf E}$, respectively.  This
means that magnetoelectricity in the system originates only from the toroidal
order. Notice that Eq. (\ref{Max3}) just expresses that the magnetoelectric
tensor  $\alpha_{ij}$  obeys $P_i= \chi_{ij}^pE_j+\alpha_{ij}H_j$ and $M_i 
= \alpha_{ij}E_j+\chi_{ij}^m H_j$, where $\chi_{ij}^p$ and $\chi_{ij}^m$ are 
dielectric and magnetic susceptibility tensors, respectively.  From the $\tau\cdot G$ 
term in Eq.~(9) it also follows that the components of the toroidal moment obey 
the proportionality: $\tau_1 \sim(\alpha_{23}-\alpha_{32})$, $\tau_2 \sim(\alpha_{31} 
-\alpha_{13})$ and $\tau_3 \sim(\alpha_{12}-\alpha_{21})$. 

An interesting relationship between ${\bf \Tau}$ and ${\bf P}_t$, ${\bf M}_t$ can be obtained considering that:
%
\begin{eqnarray}
{\bf P}_t \times {\bf M}_t & = & - ({\bf \Tau} \times {\bf H}) \times ({\bf \Tau} \times {\bf E}) \nonumber \\ & = & ({\bf \Tau} \times {\bf E})
\times ({\bf \Tau} \times {\bf H}) \nonumber \\ & = & -({\bf \Tau} \times {\bf E}) \times ({\bf H} \times {\bf \Tau}) \nonumber \\ & = &  - {\bf
\Tau} \times ({\bf E} \times {\bf H}) \times {\bf \Tau} \nonumber \\ & = & -{\bf \Tau}\times {\bf G} \times {\bf \Tau} .
\end{eqnarray}
Taking into account the general vectorial relation ${\bf A} \times {\bf B} \times {\bf A} = {\bf A} ({\bf A} \cdot {\bf B}) - \frac{1}{2} A^2
{\bf B}$, we can rewrite the preceding equation as:
\begin{equation}
{\bf P}_t \times {\bf M}_t =  \frac{1}{2}\tau^2 {\bf G} - {\bf \Tau} ({\bf G} \cdot {\bf \Tau}),
\end{equation}
which points out that ${\bf P}_t \times {\bf M}_t$ is different from zero only under an applied toroidal field ${\bf G}$.   Here $\tau=|{\bf
\Tau}|$ denotes the magnitude of toroidization.

\subsection{The toroido-caloric effect}

The equation quantifying the toroidocaloric effect under an applied electric or magnetic field is  obtained using Maxwell relations (\ref{Max1})
and (\ref{Max2}). We obtain the following isothermal changes of entropy:
%
%
%
%
%
\begin{eqnarray}
\Delta S(T, 0 \rightarrow {\bf E}) \;\;\;\;\;\;\;\;\;\;\;\;\;\;\;\;\;\;\;\;\;\;\;\;\;\;\;\;\;\;\;\;\;\;\;\;\;\;\;\;\;\;\;\;\;\;\;\;\;\;\;\;\;\;
 \nonumber \\ = \int_0^{\bf E} \left[\left(\frac{\partial {\bf P}_i}{\partial T} \right)_{{\bf E},{\bf H}} + {\bf H} \times \left(\frac{\partial {\bf \Tau}}{\partial
T} \right)_{{\bf E},{\bf H}} \right] \cdot d{\bf E}
\end{eqnarray}
and
\begin{eqnarray}
\Delta S(T, 0 \rightarrow {\bf H}) \;\;\;\;\;\;\;\;\;\;\;\;\;\;\;\;\;\;\;\;\;\;\;\;\;\;\;\;\;\;\;\;\;\;\;\;\;\;\;\;\;\;\;\;\;\;\;\;\;\;\;\;\;\;
\nonumber \\ = \int_0^{\bf H} \left[\left(\frac{\partial {\bf M}_i}{\partial T} \right)_{{\bf H},{\bf E}} + \left(\frac{\partial {\bf
\Tau}}{\partial T}\right)_{{\bf H},{\bf E}} \times {\bf E} \right] \cdot d{\bf H} .
\end{eqnarray}
The corresponding equations for the adiabatic change of temperature are:
\begin{eqnarray}
\Delta T(T_{i}, 0 \rightarrow {\bf E})
\;\;\;\;\;\;\;\;\;\;\;\;\;\;\;\;\;\;\;\;\;\;\;\;\;\;\;\;\;\;\;\;\;\;\;\;\;\;\;\;\;\;\;\;\;\;\;\;\;\;\;\;\;\;
 \nonumber \\ = -\int_0^{\bf E} \frac{T}{C}\left[\left(\frac{\partial {\bf P}_i}{\partial T} \right)_{{\bf E},{\bf H}} + {\bf H} \times \left(\frac{\partial {\bf
\Tau}}{\partial T} \right)_{{\bf E},{\bf H}} \right] \cdot d{\bf E}
\end{eqnarray}
and
\begin{eqnarray}
\Delta T(T_{i}, 0 \rightarrow {\bf H})
\;\;\;\;\;\;\;\;\;\;\;\;\;\;\;\;\;\;\;\;\;\;\;\;\;\;\;\;\;\;\;\;\;\;\;\;\;\;\;\;\;\;\;\;\;\;\;\;\;\;\;\;\;\;
 \nonumber \\ = -\int_0^{\bf H} \frac{T}{C}\left[\left(\frac{\partial {\bf M}_i}{\partial T} \right)_{{\bf H},{\bf E}} + \left(\frac{\partial {\bf \Tau}}{\partial
T}\right)_{{\bf H},{\bf E}} \times {\bf E} \right] \cdot d{\bf H} .
\end{eqnarray}
The second term in the square brackets in Eqs. (20) - (23) represents the toroidal contribution. It is interesting to notice that when applying an
electric (magnetic) field, no toroidal contribution to the caloric effect will occur if the magnetic (electric) field is zero.

%
If only toroidal order is present in the system the above equations for isothermal entropy change reduce to:
\begin{equation}
\Delta S(T, 0 \rightarrow {\bf E}) = \int_0^{\bf E} \left(\frac{\partial {\bf P}_t}{\partial T} \right)_{{\bf E},{\bf H}} \cdot d{\bf E}
\end{equation}
and
\begin{equation}
\Delta S(T, 0 \rightarrow {\bf H}) = \int_0^{\bf H} \left(\frac{\partial {\bf M}_t}{\partial T} \right)_{{\bf H},{\bf E}} \cdot d{\bf H} .
\end{equation}
One finds similar expressions for the adiabatic temperature change $\Delta T$ $(T_{i}, 0 \rightarrow {\bf E}) $ and $\Delta T$ $(T_{i}, 0 \rightarrow
{\bf H})$.

\section{Landau model for a material undergoing a ferrotoroidic transition}

We assume that the toroidization ($\tau =|{\bf \Tau}|$) is the order parameter
to describe a paratoroidic to ferrotoroidic transition and propose the
following minimal Landau model which includes a coupling between toroidization
($\tau$), magnetization ($M$) and polarization ($P$), and the presence of
toroidal ($G$), magnetic ($H$) and electric ($E$) fields:
%
%
%
\begin{eqnarray}
F(T, {\bf \Tau}, {\bf P}, {\bf M}) & = & \frac{1}{2}A_0(T-T_c^0) \tau^2 +
\frac{1}{4} C \tau^4 \nonumber \\ & + & \frac{1}{2} \chi_p^{-1} P^2 +
\frac{1}{2}
\chi_m^{-1} M^2 \nonumber \\
& + &  \kappa {\bf \Tau} \cdot ({\bf P} \times {\bf M}) - {\bf G} \cdot {\bf \Tau} \nonumber \\ & - & {\bf H} \cdot {\bf M} -{\bf E} \cdot {\bf
P}. \label{free-energy}
\end{eqnarray}
Note that the term $\kappa$ is the lowest order symmetry allowed term
(satisfying space and time reversal symmetry) which provides the coupling
between toroidization, polarization and magnetization (see
\cite{Ederer2007}). Here the inverse toroidic susceptibility $\chi_t^{-1} =
A_0(T-T_c^0)$, where $A_0$ is the ``toroidic stiffness", $T_c^0$ is the
transition temperature and $C>0$ is the nonlinear toroidic coefficient. In
particular, $\chi_t = \partial{\bf P}/\partial{\bf H} = \partial{\bf
M}/\partial{\bf E}$.  We also note that a Landau theory of ferrotoroidic
transitions in boracites \cite{sannikov} and Ba$_2$CoGe$_2$O$_7$ (BCG)
\cite{toledano} has been considered previously. However, these studies did not
consider any caloric effects.

Minimization with respect to polarization and magnetization gives:
\begin{equation}
\frac{\partial F}{\partial {\bf P}} = \chi_p^{-1} {\bf P} - {\bf E} + \kappa ({\bf M} \times {\bf \Tau}) = 0
\end{equation}
and
\begin{equation}
\frac{\partial F}{\partial {\bf M}} = \chi_m^{-1} {\bf M} - {\bf H} + \kappa ({\bf \Tau} \times {\bf P}) = 0 .
\end{equation}
We now solve these two equations assuming (for simplicity) that ${\bf E} =
(E,0,0)$ and ${\bf H} = (0,H,0)$ and therefore ${\bf G} = (0,0, EH)$ and ${\bf
\Tau} = (0,0,\tau)$. For ${\bf P}$ [$= (P,0,0)$] and ${\bf M}$ [$ = (0,M,0)$]
we obtain:
\begin{equation}
P = \chi_p E - \kappa \chi_p \chi_m H \tau + O(\tau^2) \simeq \chi_p E - \alpha H  \label{eqP}
\end{equation}
and
\begin{equation}
M = \chi_m H - \kappa \chi_p \chi _m E \tau + O(\tau^2) \simeq \chi_m H - \alpha E  , \label{eqM}
\end{equation}
%
where in the above two equations we have neglected the nonlinear magnetoelectric effects.
The magnetoelectric coefficient $\alpha = \kappa\chi_p\chi_m\tau$ is a
quadrilinear product of electric susceptibility ($\chi_p=\partial{\bf
P}/\partial{\bf E}$), magnetic susceptibility ($\chi_m = \partial{\bf
M}/\partial{\bf H}$), the coupling constant $\kappa$ and the toroidization
$\tau$.  Thus, either for $\kappa = 0$ or $\tau=0$ there is no magnetoelectric
effect. Substitution of $P$ (\ref{eqP}) and $M$ (\ref{eqM}) in the free energy
(\ref{free-energy}) gives the following general type of effective free energy:
%
%
\begin{eqnarray}
F_{e} & = & F_0(E, H) \nonumber \\ & + & \frac{1}{2} A \tau^2 + \frac{1}{3} B \tau^3 + \frac{1}{4} C \tau^4 + \lambda \tau ,
\label{eff-free-energy}
\end{eqnarray}
with
\begin{eqnarray}
F_0 & = & -\frac{1}{2} \left(\chi_p E^2 + \chi_m H^2 \right) , \\
A & = & A_0(T - T_c^0) - \kappa^2 \chi_p \chi_m [\chi_m H^2 + \chi_p E^2]
\nonumber \\ & = & A_0(T - T_c) , \\
T_c & = & T_c^0 + \frac{\kappa^2}{A_0} \chi_p \chi_m [\chi_m H^2 + \chi_p E^2] ,
\label{Tc}\\
B & = & 3\kappa^3 \chi_m^2 \chi_p^2 EH , \label{B}\\
\lambda & = & (\kappa  \chi_m  \chi_p - 1) EH .
\end{eqnarray}
This corresponds to the free energy of a system subjected to an effective
external field $\lambda$, proportional to the toroidal field $G=EH$.  When
$G = 0$, and therefore from (\ref{B}) $B=0$, the free
energy  (\ref{eff-free-energy}) describes a paratoroidal-to-ferrotoroidal
second-order phase transition whereas under the application of a toroidal
field $G \neq 0$ (and $B \neq 0$), the transition becomes a first-order one. Actually,
the physics contained  in (\ref{eff-free-energy}) is very rich since, in addition to
$\lambda$, the cubic $B$ coefficient also depends on the toroidal field $G$ and
the linear $A(T)$ term explicitly depends on the $E$ and $H$ fields.  In other words,
$\lambda$, $A(T)$ and $B$ are not independent. This leads to a competition
between  $B$ and $\lambda$ depending on the value of the coupling constant
$\kappa$ (i.e. the choice of the ferrotoroidic material) and to a nonlinear
Clausius-Clapeyron equation.
It is worth pointing out that the addition of nonlinear terms in eqs. (\ref{eqP})
and (\ref{eqM}) would lead to higher order terms in the expansion
(\ref{eff-free-energy}) that go beyond the minimal model. However, within the
spirit of the Landau Theory, we expect
such terms are not essential.

For some purposes, it may be convenient to rescale the free energy
according to the definitions proposed in \cite{Sanati2003}:
\begin{eqnarray}
a_0 & = & \frac{A_0 C}{B^2} , \nonumber \\
h & = & \frac{\lambda C^2}{|B|^3} , \nonumber \\
\tau_r & = & \frac{C}{|B|} \tau , \nonumber \\
f_r & = & \frac{F_{eff}}{B^4}C^3 ,
\label{scaling}
\end{eqnarray}
which yields the following rescaled free energy, valid for $B \neq 0$ only:
\begin{equation}
{f_r} = f_0 + \frac{1}{2}\; a \; {\tau_r}^2 + \frac{1}{3} \; {\tau_r}^3 +
\frac{1}{4}\; {\tau_r}^4 + h {\tau_r}.
\label{rescaled}
\end{equation}
As usual, the liner coefficient $a$ is temperature dependent $a = a_0 (T -
{T}_c)$ and provides the temperature scale, while $h$ provides the scale of the
external effective field.

\subsection{Phase diagram}
\begin{figure}[ht]
\begin{center}
\epsfig{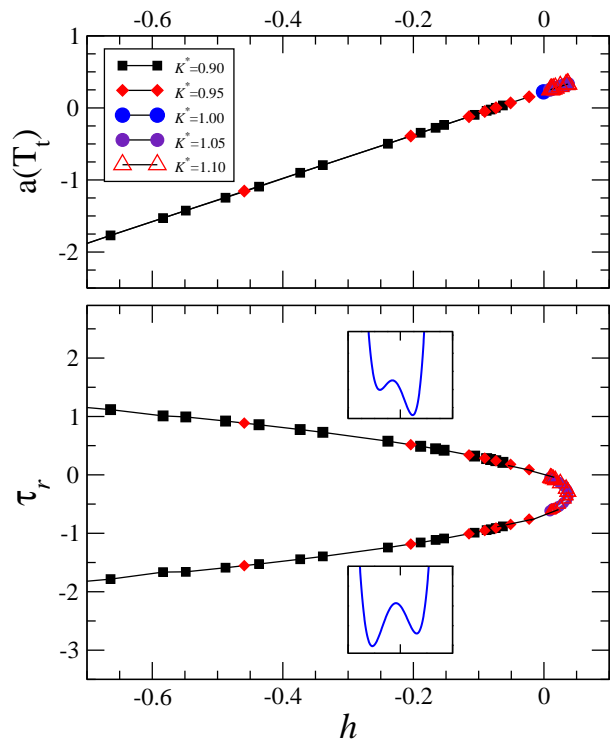}
\end{center}
\caption{(color online). Rescaled phase diagram as a function of the field $h$ for the
  temperature coefficient $a(T)$ (upper panel) and the order parameter $\tau_r$
  (lower panel). Different symbols
  denote results obtained for different values of the effective coupling constant
  $\kappa^*=\kappa \chi_p \chi_m$.} \label{FIG1}
\end{figure}

Figure \ref{FIG1} shows the phase diagram for the rescaled variables
(\ref{scaling}) defined above. In the upper panel we have plotted the behavior
of $h$ vs.  $a(T)$ which gives the coexistence line.  The region above the
line is paratoroidic whereas below it is ferrotoroidic. The coexistence line
ends in a critical point that can be calculated from  the condition:
\begin{equation}
-\frac{\partial h}{\partial \tau} = a + 2 \tau + 3 \tau^{2} = 0 .
\end{equation}
One obtains the critical point is located at ($h_c$,$a_c$)=($\frac{1}{27}$,$\frac{1}{3}$).
Notice that the field is $h \leq 0$
for $\kappa^* \leq 1$ whereas for  $\kappa^* > 1$
we have $h>0$ but $h<h_c = \frac{1}{27}$. The existence of
this critical point is also revealed from the turning point in the behavior of
$\tau_r$ vs. $h$ shown in the lower
panel. Below the critical field ($h < h_c$), there are two possible values of $\tau_r$
related to the two possible wells in the free energy, as it is schematically
illustrated in the insets. Different symbols
denote that the results have been obtained for different values of the effective coupling parameter
$\kappa^* = \kappa \chi_m \chi_p$ by solving numerically the model (\ref{eff-free-energy}). The
subsequent application of scaling relations defined in (\ref{scaling}) leads to the
curves shown in Fig. \ref{FIG1}.

\begin{figure}[ht]
\begin{center}
\epsfig{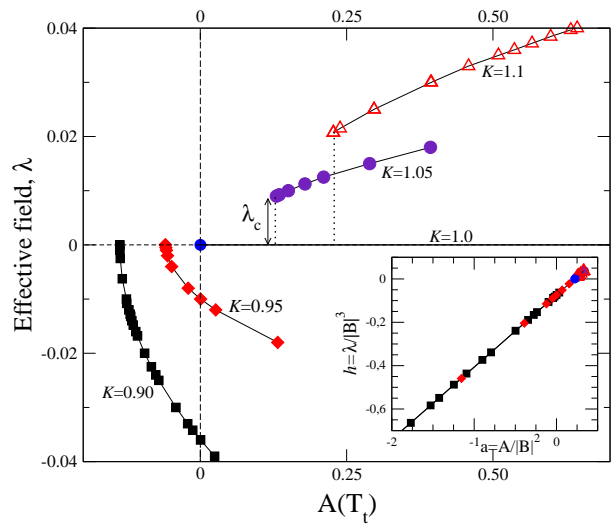}
\end{center}
\caption{(color online). Phase diagram showing the effective field $\lambda$ as a function of
  the transition temperature $A(T_t)=T_t-T_c$ for representative values of
  $\kappa > 1$, $\kappa = 1$ and $\kappa < 1$. The inset shows the coexistence
  line in  rescaled
  variables. The (blue) dot on the $\lambda = 0$ line corresponds to the case
  of $B=0$.}
\label{FIG2}
\end{figure}

Actually, the unscaled free-energy (\ref{eff-free-energy}) is more suitable for studying
independently the effect of the external toroidal
field $G$ and the material dependent coupling parameter $\kappa$. For
convenience, we set $\chi_p = 1$, $\chi_m
=  1$, $C = 1$, $T_c^0 = 1$ and $A_0 = 1$, which renders the following
dependence for the model parameters:
\begin{eqnarray}
B & = & 3 \kappa^3 EH = 3\kappa^3G ,\\
\lambda & = & (\kappa - 1) EH = (\kappa - 1)G , \\
A(T) & = & T - T_c , \\
T_c & = & 1 +  \kappa^2 (E^2 + H^2) .
\end{eqnarray}
The only free parameter is the coupling constant $\kappa$. From now on
we will restrict to values of $\kappa > 0$ and consequently to $B \geq 0$.
Under these conditions,  different situations can be considered:
\begin{itemize}
\item $B > 0$ and $\lambda = 0$, corresponds to $\kappa=1$. In the low
  temperature regime the solution corresponds to a minimum located at $\tau < 0$.
\item $B > 0$ and $\lambda > 0$, corresponds to $\kappa > 1$.
In this case the low temperature minimum also occurs at $\tau < 0$.
\item $B > 0$ and $\lambda < 0$, corresponds to $0 < \kappa < 1$.
In that case, competition between $\tau < 0$ and $\tau > 0$ occurs: $\lambda < 0$
favors the minimum to occur at $\tau > 0$ while $B > 0$ favors the minimum to occur
at $\tau < 0$.
\end{itemize}
As a reference case,  we also consider the possibility of $G=0$. In that situation,
\begin{itemize}
\item $B = 0$ and $\lambda = 0$ for all values of $\kappa$.
The transition is continuous and occurs at $T_c^0 =1$. In the low
temperature phase the free energy shows two symmetric minima at $\pm \tau_0$.
\end{itemize}
The corresponding phase diagram obtained from the unscaled free energy
(\ref{eff-free-energy}) is depicted in Fig. \ref{FIG2}. It shows the behavior
of the effective field $\lambda$ as a function of the transition temperature
$T_t$ (or $A(T_t)= T_t-T_c$) for the three different regimes of the coupling
parameter mentioned above; namely $\kappa>1$, $\kappa=1$ and $\kappa<1$. (i)
For $\kappa >1$ (above) the transition exists for values of $\lambda$ above
the critical field $\lambda_c$($\kappa$) $>0$, which in turn depends on the
value of $\kappa$. Moreover, $T_t$ increases with $\lambda$ as it can be seen
from the representative curves obtained for $\kappa=1.05$ and $1.10$.  (ii)
For $\kappa = 1$ the effective field is $\lambda = 0$ and the coexistence line
is horizontal starting at the critical point (at the origin) corresponding to
the limiting case of $G=0$ ($B=0$) and denoted by a blue dot. (iii) Finally
for  $\kappa < 1$, one has $\lambda < 0$ and the transition exists for  values
of $\lambda$ arbitrarily close to zero at  temperatures $T_t < T_c$.  For
increasing values of $G$, $\lambda$ decreases while $A(T_t)$ increases;   thus
eventually the latter may become positive.  We stress that all curves
correspond to discontinuous phase transitions except the point located at the
origin that corresponds  to $B=0$ and therefore to a continuous phase
transition.

\begin{figure}[ht]
\begin{center}
\epsfig{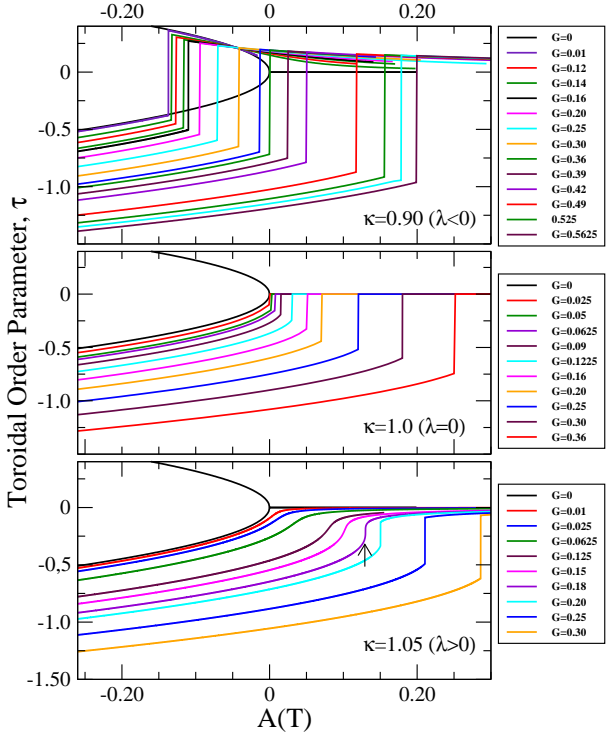}
\end{center}
\caption{(color online). Toroidal order parameter as a function of temperature for $\kappa = 0.90$, $\kappa = 1$ and $\kappa = 1.05$ and selected values of the
applied toroidal field $G$. The arrow in the lower panel indicates the occurrence
  of the discontinuous transition.}
\label{FIG3}
\end{figure}

Figure \ref{FIG3} shows the temperature behavior of the toroidal order parameter $\tau$
for three representative values
of the coupling parameter namely $\kappa=0.90,1.0,1.05$ and different values of the
toroidal field $G=EH$ as indicated in the lateral panels.
\begin{figure}[ht]
\begin{center}
\epsfig{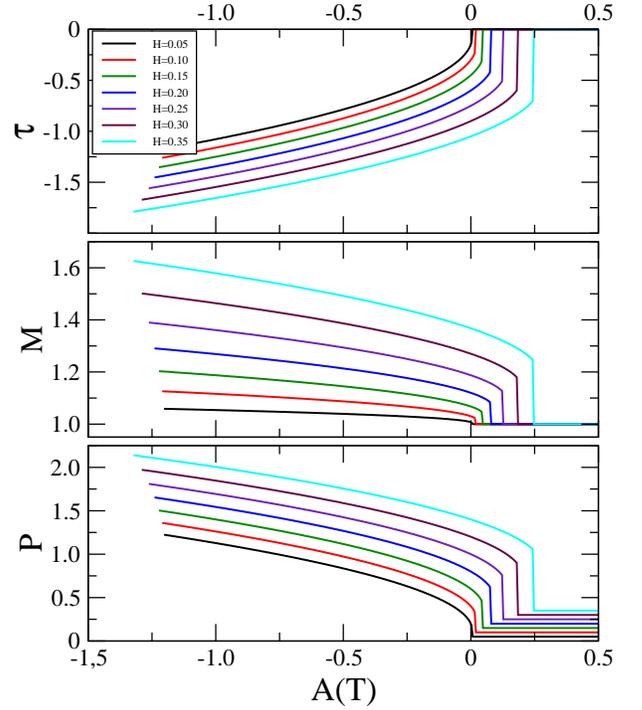}
\end{center}
\caption{(color online). Toroidal order parameter ($\tau$), magnetization ($M$)  and
  polarization ($P$)  as a
  function of temperature $A(T)=T-T_{c}$ for $\kappa = 1.0$. The toroidal field is modified
  by changing the magnetic field $H$ but keeping constant the value of the
  electric field $E=1.0$.}
\label{FIG4}
\end{figure}
The case $\kappa = 1.05$ (lower panel) nicely illustrates the effect of the
field on the ferrotoroidal transition in the region of $\kappa > 1$.
For zero-field ($G=0$), the transition is
continuous and, indeed, occurs at $T=T_c^0$ = 1 (symmetric curve with
double branch). By increasing the field ($G>0$), the transition first disappears (continuous
cross-over, no singularity, from $\tau \neq 0$ to
$\tau = 0$) and subsequently for higher values of the field a first-order
transition occurs at a temperature $T_t > T_c^0$ revealed by a discontinuous
jump in $\tau$ (indicated by an arrow), whose magnitude increases with $G$.
In the case of $\kappa=1$ (middle
panel), the transition is discontinuous and exists for every value of $G \neq
0$, although it is continuous for $G=0$. In the case $\kappa < 1$ as soon as
the field is applied a first-order transition from $\tau<0$ to $\tau>0$ occurs
with cooling. For low values of $G$, the transition  takes place at $T_t < T_c$
but with increasing values of $G$, $T_t$ also increases and the transition takes
places above $T_c$. This behavior results from the competition between $B$
and $\lambda$.

The temperature behavior of the polarization $P$ and the magnetization $M$ can then be
obtained from the expressions (\ref{eqP}) and (\ref{eqM}), respectively. The
results are shown in Fig. \ref{FIG4} for the case of $\kappa = 1.0$ and
different values of $G=EH$ ($E=1.0$).

\begin{figure}[ht]
\begin{center}
\epsfig{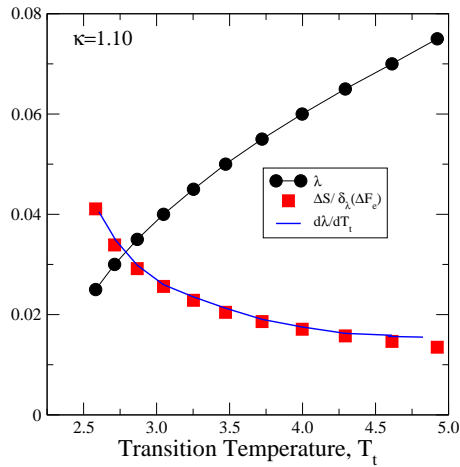}
\end{center}
\caption{(color online). Clausius-Clapeyron equation. We have plotted the coexistence line
  (black dots) for $\kappa =1.10$ obtained at constant $E=1.0$. Squares in red denote
  the results of computing the right hand side of Eq. (50) from the
  discontinuities of $\tau$  and $S$  at the transition point. Finally, the continuous
  line (blue) is the numrical derivative of the coexistence line.}
\label{FIG5}
\end{figure}
\subsection{Nonlinear Clausius-Clapeyron Equation}

Since the transition predicted by the model is discontinuous, it can be
characterized by means of the corresponding Clausius-Clapeyron equation.
Such equation relates the slope of the coexistence curve (Fig. \ref{FIG2}) to the
magnitude of the discontinuities in the order parameter  and entropy
at the transition temperature. We recall that in the present model, the
harmonic ($A(T)$) and cubic ($B$) coefficients and the effective
field ($\lambda$) are not independent. As noted below, this will give rise to a Clausius-Clapeyron
equation which turns out to be nonlinear in the order parameter discontinuity.

Indeed, at the transition, one can write:
\begin{equation}
d{F_{e}}^{(I)} =  d{F_{e}}^{(II)} ,
\label{coexistence}
\end{equation}
where (I) and (II) are the para- and ferrotoroidal phases respectively.
For each phase, the derivative of the effective free energy can be written as:
\begin{eqnarray}
dF_{e} &=& \left( \frac{\partial F_{e}}{\partial T} \right)_{\lambda} dT + \left(\frac{\partial
  F_{e}}{\partial \lambda}\right)_{T} d\lambda  ,
\end{eqnarray}
where $F_e(T,\lambda)$ is the thermodynamic free energy given by the expression
(\ref{eff-free-energy}) with $\tau=\tau(T)$ in equilibrium. The partial derivatives are:
\begin{eqnarray}
\left(\frac{\partial F_{e}}{\partial T}\right)_{\lambda} &=& -S =
-\frac{A_0}{2} \tau^{2} \label{partial1_F_e} , \\
\left(\frac{\partial F_{e}}{\partial \lambda}\right)_{T} &=& \frac{1}{E (\kappa
  -1)}\left(\frac{\partial F_{e}}{\partial H} \right )_{T} ,
\label{partial2_F_e}
\end{eqnarray}
where $S$ is the entropy and the derivative with respect to the field $\lambda$ is
taken at constant $E$.
\begin{equation}
\left(\frac{\partial F_{e}}{\partial H}\right)_{T} = -H\left(1+A_{0}
{\kappa}^{2} {\tau}^{2}\right) +  E \left({\kappa}^{3}
{\tau}^{3} +  {\tau} (\kappa-1) \right)
\label{partial_Fe_H}  .
\end{equation}
Taking into account (\ref{partial1_F_e}), (\ref{partial2_F_e})  and
(\ref{partial_Fe_H}), condition (\ref{coexistence}) reads:
\begin{eqnarray}
dT(S_{II}-S_{I}) &=&  \partial_{\lambda}(\Delta F_{e}) d\lambda =  \\ &=&
\frac{d \lambda}{E (\kappa -1)}  \left [ {\left(\frac{\partial F_{e}}{\partial
        H} \right)}^{(II)}_{T} - {\left(\frac{\partial F_{e}}{\partial
        H} \right)}^{(I)}_{T} \right ] \nonumber
\end{eqnarray}
with
\begin{eqnarray}
 & &{\left(\frac{\partial F_{e}}{\partial
        H} \right)}^{(II)}_{T} - {\left(\frac{\partial F_{e}}{\partial
        H} \right)}^{(I)}_{T} = \\
&=& \left [({\tau}^{3}_{(II)}-
    {\tau}^{3}_{(I)})  E - ({\tau}^{2}_{(II)}-
    {\tau}^{2}_{(I)})H \right] {\kappa}^{3} +  \nonumber \\ &+& (\tau_{(II)} - \tau_{(I)}) (\kappa-1) E .
\nonumber
\end{eqnarray}

Finally, the corresponding Clausius-Clapeyron equation takes the form:

\begin{eqnarray}
\frac{d \lambda}{dT} = \frac{\Delta S}{\partial_{\lambda} (\Delta F_{e})} =
\frac{(S_{II}-S_{I}) E (\kappa-1)}{{\left(\frac{\partial F_{e}}{\partial
        H} \right)}^{(II)}_{T} - {\left(\frac{\partial F_{e}}{\partial
        H} \right)}^{(I)}_{T}}  .
\label{CC}
\end{eqnarray}
Notice that for $\kappa=1$, the slope of the coexistence curve is zero although
there exists a jump in both $\tau$ and $S$ at the transition point. In Fig.
\ref{FIG5} we have plotted the coexistence line for $\kappa =1.10$ (black
dots) and its numerical derivative (continuous blue line). The squares (full red) are the
results of computing the right hand side of Eq. (\ref{CC}) from the magnitude of
the discontinuity of $\tau$ and $S$ at the transition temperature. The
agreement is very good.

\subsection{Toroidocaloric effect}
\begin{figure}[ht]
\begin{center}
\epsfig{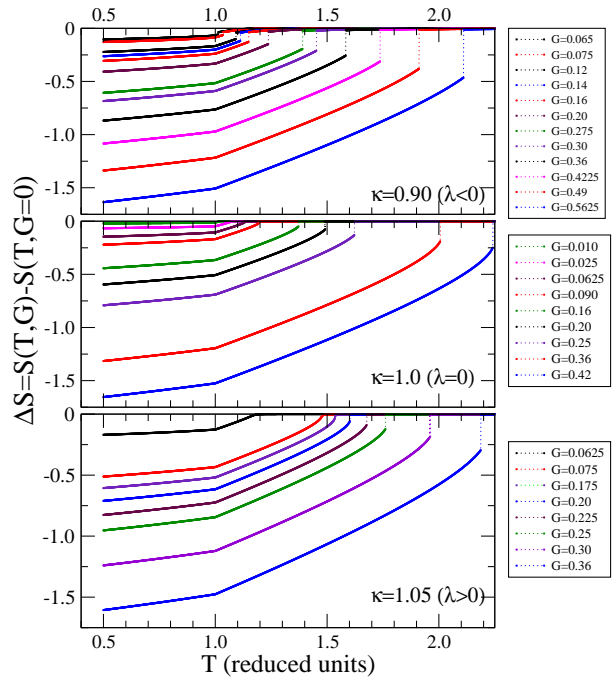}
\end{center}
\caption{(color online). Toroidocaloric effect as a function of temperature and selected
  values of the maximum applied toroidal field, $\kappa = 0.90, 1.0$ and $1.05$.}
\label{FIG6}.
\end{figure}

The toroidocaloric effect can be computed as:
\begin{eqnarray}
S(T, G=EH) - S(T, G = 0) =  \int_0^{G=EH} \left(\frac{\partial \tau}{\partial T} \right)_G d G  \nonumber
\\ =  \frac{1}{2} A_0 [\tau^2(T, G=0) - \tau^2(T, G =EH)]  . \;\;\;\;\;\;\;\;\;\;\;\;\;\;
\end{eqnarray}
This is an important equation for the isothermal toroidal field-induced entropy change which is analogous to the expression found long ago
for the adiabatic electric field-induced temperature change in the case of the electrocaloric effect \cite{strukov1967}. The isothermal entropy
change ($\Delta S$) as a function of temperature for various values of applied toroidal field ($G=EH$) has been plotted in Fig. \ref{FIG6} for the
same three representative values of $\kappa$ as in Fig. \ref{FIG3}. As it can be observed, with increasing field value the jump in $\Delta S$
increases.
%
%

It is interesting to relate this entropy change characterizing the toroidocaloric effect with the corresponding
changes giving the magnetocaloric and electrocaloric effects. This can be done using Eqs. (\ref{eqP}) and
(\ref{eqM}), a straightforward calculation gives:
\begin{eqnarray}
\Delta S(T, 0 \rightarrow G =EH) \;\;\;\;\;\;\;\;\;\;\;\;\;\;\;\;\;\;\;\;\;\;\;\;\;\;\;\;\;\;\;\;\;\;\;\;\;\;\;\;\;\;\;\;\;\;\;\; \nonumber \\  =
\frac{1}{\kappa \chi_p \chi_m} [\Delta S_P(T, H, 0 \rightarrow E) + \Delta S_M(T, E, 0 \rightarrow H)] \nonumber \\,
\label{entropies}
\end{eqnarray}
where $\Delta S_P$ and $\Delta S_M$ are the entropy changes giving the electrocaloric and the
magnetocaloric effects at constant magnetic and electric fields, respectively.

To apply our results to a specific material we need to obtain parameters (e.g. $\kappa$) for LiCo(PO$_4$)$_3$
using the data from second harmonic generation \cite{vanaken} and other thermodynamic measurements, in
particular the toroidic susceptibility.  The latter is related to the antisymmetric part of the magnetoelectric tensor 
as discussed in \cite{kadom} for BiFeO$_3$.  It is interesting to point out that Eq. (\ref{entropies}) suggests that 
if the electric and magnetic susceptibilities are known, $\kappa$ could be estimated from measurements of the 
entropy changes $\Delta S$, $\Delta S_P$ and $\Delta S_M$.

\section{Conclusions}
With the discovery of the fourth kind of primary  ferroic materials, namely the ferrotoroidics \cite{schmid, spaldin}, it is important to
understand the equilibrium thermodynamic properties of such materials. We provided a basic framework to calculate the toroidocaloric effect based
on a Landau free energy. Our main finding is the isothermal change in entropy as a function of the applied toroidal field for different values of
coupling between toroidization, polarization and magnetization.  The fact that the application of a toroidal field modifies the order of the
transition from continuous to first-order is very informative regarding caloric effects since larger changes of entropy are expected to be induced
thanks to the large entropy content associated with the latent heat of a first-order transition. However, when dealing with real materials, the
existence of energy losses arising from hysteresis and domain wall effects \cite{prosandeev2008b,review} (which are intrinsic to first-order
transition and are expected to reduce the magnetocaloric efficiency) should be taken into account. This important aspect has not been considered
in the present paper since strict equilibrium situations are assumed. In any case, our predictions should be observable in experiments on
materials such as LiCo(Po$_4$)$_3$, Ba$_2$CoGe$_2$O$_7$ (BCG), MnTiO$_3$, MnPS$_3$ and some boracites. Below 21.8 K neutron diffraction
measurements in LiCo(Po$_4$)$_3$ indicate simultaneous presence of ferrotoroidic and antiferromagnetic (AFM) order as a result of Co$^{2+}$ ion
ordering \cite{vanaken}.

Similarly,
in BCG there is a transition at 6.7 K below which there is a coexistence of ferrotoroidic and AFM order again resulting from single ion effects
\cite{toledano}.  In the ilmenite structure MnTiO$_3$ there is an antiferromagnetic ordering below 63.5 K.  At cryogenic temperatures it exhibits
ferrotoroidic behavior above 6 T magnetic field due to spin flopping \cite{tokura}. These results confirm the existence of a ferrotoroidic phase
in a number of materials, however, the measurement of the toroidal moment as a function of temperature and toroidal field has not been undertaken.
Therefore, at the present stage it is not possible to contrast our predictions with experimental results. We expect, however, that our results including the nonlinear Clausius-Clapeyron relation will be
a motivation for experimentalist to undertake experiments aimed at characterizing the thermodynamic behavior of ferrotoroidic materials in the
vicinity of the phase transition.

Using the Landau free energy we can also obtain the profiles of ferrotoroidic domain walls by including symmetry allowed gradient term $(\nabla
\tau)^2$ \cite{kopaev}, i.e. the Ginzburg term. Such domain walls have been observed in LiCo(Po$_4$)$_3$ using optical second harmonic generation techniques
\cite{vanaken}. With doping induced disorder in such materials we expect that novel phases such as toroidic tweed and toroidic glass should also
exist and remain to be observed experimentally. With symmetry allowed coupling of strain to toroidization, if we apply stress to such a crystal we
expect toroidoelastic effects, i.e. a change in toroidization with hydrostatic pressure or shear.  These important topics will be
explored in near future.

\begin{acknowledgments}
We acknowledge COE Program, Japan for supporting the visit to Osaka University of AS and AP where this work was initiated. This work was supported
in part by the U.S. Department of Energy and CICyT (Spain), Project MAT2010-15114.
\end{acknowledgments}

\end{document}